\documentclass[11pt]{article}

\usepackage[a4paper, total={6.5in, 9in}]{geometry}
\usepackage{graphicx}
\usepackage[usenames, dvipsnames]{color}
\usepackage{shapepar}
\usepackage{picinpar}
\usepackage{amsmath}
\usepackage[square]{natbib} 
\usepackage{marvosym}
\usepackage{eurosym}
\usepackage{lineno}

\usepackage{makeidx}
\makeindex

\begin{document}

\begin{center}
{\Large \bf 2. Origin of Ganymede and the Galilean Moons} \\
{\Large Yuhito Shibaike and Yann Alibert} \\
\end{center}

\section{Introduction}
The four large moons orbiting around Jupiter, found by Galileo Galilei in 1610, are called Galilean moons\index{Galilean moons}. Ganymede is the largest of them, and the largest moon in our solar system. The Galilean system can be seen as a miniature version of planet systems and its origin is potentially also similar to that of planets. We start this chapter by a comparison between the processes of satellite and planet formation.

Ganymede and the Galilean moons are believed to have formed in a gas disc, like planets formed in protoplanetary discs (PPDs). At the end of its formation, Jupiter accreted gas from the circum-stellar disc (CSD) and, as a byproduct, a gas disc formed around the gas planet, which is called the "circum-Jovian disc" (CJD). The formation processes of the large moons are similar to those of planets; the seeds of the moons grow large by accreting "satellitesimals\index{Satellitesimal}" (similar to planetesimals\index{Planetesimals} in the case of planets, a few kilometers in size) and/or pebbles\index{Pebbles} (centimeter to meter size). The moons also migrate in the CJD, and the orbits get into resonances. Due to the similarity, the physics and technical methods involved in planet formation have also been applied to the formation of the large moons in many ways. However, there are mainly three differences between them.

\begin{itemize}
\item {\bf Small scale} One of the most important difference is the scale. The semi-major axis of Ganymede is about $10^{6}~{\rm km}$, which is more than 100 times smaller than that of Earth ($1~{\rm au}=1.5\times10^{8}~{\rm km}$). The orbital period of Ganymede is 7 days, 50 times shorter than that of Earth, suggesting that the timescale of the physical processes of satellite formation is roughly 10 to 100 times shorter than that of the planet formation. The mass ratio between the Sun and Jupiter is about $10^{3}$, and the temperature of the CJD and CSD is not so different, so the aspect ratio of the CJD is about three times larger than that of the CSD. Due to the short typical length scale of the CJD, the magnetic diffusion is also much faster (i.e., the magnetic Reynolds number is small) and it prevents magnetorotational instability\index{Magnetorotational instability} (MRI) in broad regions of the disc \citep[e.g.,][]{fuj14,tur14} (See also Section \ref{magneticfields}).
\item {\bf Open system} The CJD is an open system unlike PPDs. In modern views of satellite formation, gas and solids are continuously supplied to the CJD from the CSD and flow into Jupiter. As a result, the continuous inflows of gas and solids dominate the formation of the large moons. This fact is interesting because the timescale of the inflow must be consistent with that of the CSD ($\sim3~{\rm Myr}$), which is much longer than the Kepler timescale inside the CJD, suggesting that the satellite formation is a quasi-steady state phenomenon that repeats over and over. Also, the difference of the timescales makes numerical calculations through the whole satellite formation processes tough. If there are special structures which prevent the mass flux from the CJD to Jupiter, the system loses the balance between the inflow and outflow and becomes unsteady. Especially an inner cavity of the CJD\index{Inner cavity of the CJD} created by the magnetic field of young Jupiter can solve one of the most severe problems of the satellite formation, a rapid loss of (proto-)satellites from the CJD by migration (See Section \ref{magneticfields}). The effects from the Sun are also unique to the satellite formation processes. The Hill sphere\index{Hill sphere} of Jupiter is distorted into a lemon-like shape and spiral shocks are excited by the Sun in the CJD \citep[e.g.,][]{zhu16} (see Section \ref{othereffects}). However, it is generally considered that the effects from the Sun onto Ganymede and the other Galilean moons are limited because their orbits are close to Jupiter enough.
\item {\bf Absence of generality} Although the goal of this chapter is reviewing the origin of Ganymede and the Galilean moons, comparisons to other moon systems and discussion of general satellite formation are important to understand it. Indeed, in the case of planet formation, recent observations of exoplanets and PPDs brought huge knowledge to the understanding of the planet formation theory. Unfortunately, in the case of satellite formation, there are only two large moon systems around Jupiter and Saturn in our solar system and no analogs have been found in exoplanet systems. Therefore, the discussion of the satellite formation is not general yet but limited to unique and historical one about the Jovian and Saturnian systems, which is similar to the situation of the planet formation until 1995 when the first exoplanet was discovered \citep{may95}.
\end{itemize}
In this chapter, we focus on the unique history of Ganymede and the Galilean moons, but the discussion can be adapted to the formation of Titan, a large moon of Saturn except for Section \ref{constraints} in which we discuss the constraints from the observations of the Galilean moons. It is also considered that the Saturnian mid-sized moons formed from the materials of the ancient massive ring spreading beyond the Roche radius \citep{cri12}. However, it should be difficult to form Titan by the scenario due to its long orbital period and heavy mass. The mass distribution of the Galileans moons (almost uniform and not depending on their orbits) is also difficult to reproduce by the scenario \citep{cri12}. The origin of the moons around Uranus and Neptune is still controversial: in-situ formation in their circum-planetary discs\index{Circum-planetary discs} (CPDs), spreading of the materials of their ancient rings, giant impacts, or captures of bodies after the formation of the planets \citep[e.g.,][]{ste84,agn06,can06,cri12,szu18}.

We also note that there is a candidate of exomoon\index{Exomoons}, but it is a huge Neptune-size moon around Kepler-1625b. It is considered that the satellite was captured by the planet after its formation phase even if the body is actually an exomoon \citep{tea18}. Dust continuum emission from a CPD around an exoplanet, PDS~70~c, has been found \citep{ise19,ben21,fas25,dom25}, and the spectral energy distribution (SED) of PDS~70~b (and c) is best fit with the model assuming the existence of its CPD \citep{chr19,sto20,wan21,chr24}. However, the estimated mass of the gas accreting planets is $\sim10~M_{\rm J}$ \citep{kep18,mul18,haf19,aoy19,has20,shi24}, where $M_{\rm J}$ is Jupiter mass, suggesting the CPD is much heavier than the CJD. Thus, it is dubious that the observed characteristics of the disc can be adapted directly to the discussion of this chapter.

\section{The circum-Jovian gas disc}
During the gas accretion of gas giants, small gas discs (CPDs\index{Circum-planetary discs}) form around the planets, because the accreting gas has some angular momentum. Here, we focus on the Jovian disc (CJD)\index{Circum-Jovian disc}, which is considered to be the birthplace of Ganymede and the other Galilean moons. First, we review the two classical models of the discs, ``minimum mass'' and ``gas-starved disc'' models in Section \ref{classical} \citep{lun82,can02}. We then explain more modern views of the discs based on the results of hydrodynamic simulations in Section \ref{modern}.

\subsection{Classical models of the CJD}
\label{classical}
The formation of the large moons around Jupiter and Saturn has been investigated a lot, and many CJD/CPD models have been proposed. First, we show the two classical models, which will be the basis of the following discussion of the modern view of the CJD and satellite formation.

\subsubsection{Minimum mass model}
\label{MMSD}
\citet{lun82} proposed ``minimum mass'' model\index{Minimum mass model}, which is based on the same concept as the ``minimum mass solar nebular'' model in the planet formation \citep{hay81}. In this model, the CJD is static and isolated from the CSD, and its solid mass is equal to the total mass of the Galilean moons, $M_{\rm T}\sim10^{-4}~M_{\rm J}$. Therefore, the total mass of the disc is about $1\%$ of Jupiter when the dust-to-gas ratio is equal to the solar composition, $0.01$. Some previous works showed that the Jovian (and Saturnian) moons can form in the minimum mass discs \citep{mos03a,mos03b,est09,mos10}. This model assumes a low-density region outside the centrifugal radius, $r_{\rm c}\equiv j^{2}/(GM_{\rm cp})=15~R_{\rm J}$, where $j$, $G$, and $M_{\rm cp}$ are the specific angular momentum of the inflowing gas during the formation of the disc, the gravitational constant, and the mass of the central planet, respectively. This assumption means that, when the disc forms, the gas flows inside $r_{\rm c}$ onto the disc, in other words, the outer edge of the gas inflow region is $r_{\rm out}=r_{\rm c}$. Thanks to this assumption, Callisto can avoid its internal differentiation by reducing the accretion of solid materials (see also Section \ref{internal}). Here, the specific angular momentum is assumed as $j=1/4\Omega_{\rm K,cp}R_{\rm H}^{2}$, where $\Omega_{\rm K,cp}$ and $R_{\rm H}$ are the Kepler angular velocity of the central planet rotating around the star and the Hill radius\index{Hill radius}, respectively, and the gravitational effect is ignored. They are $\Omega_{\rm K,cp}=\sqrt{GM_{\rm star}/a_{\rm cp}^{3}}$ and $R_{\rm H}=(M_{\rm cp}/M_{\rm star})^{1/3}a_{\rm cp}$, where $M_{\rm star}$ and $a_{\rm cp}$ are the mass of the central star and the semimajor axis of the planet, respectively. Here, $M_{\rm cp}$, $M_{\rm star}$, and $a_{\rm cp}$ are set as one Jupiter mass, one Solar mass, and the semimejor axis of current Jupiter, respectively. The concept that the formation process of Callisto is different from that of the other Galilean moons is inherited to the subsequent newer formation scenarios (see Section \ref{formation}) \citep{shi19,bat20}.

However, hydrodynamic simulations\index{Hydrodynamic simulations of the CJD} both in 2D and 3D suggest that the gas accretion\index{Gas accretion of planets} onto CPDs continues even when the gas gap structures\index{Gap structure of gas} open around the discs \citep[e.g.,][]{lub99,kle01}. The depth of the gaps created by gas giants are indeed predicted to be shallower than that of the traditional predictions; the gas surface density in the gaps is about 100 times smaller than that of the unperturbed regions \citep[e.g.,][]{kan15b}. It has also been considered that it is difficult to avoid rapid inward migration of the moons induced by the interaction with the massive disc \citep{mig16}.

\subsubsection{Gas-starved disc model}
\label{gas-starved}
\citet{can02} proposed ``gas-starved disk'' model\index{Gas-starved disc model}, which has a smaller gas disc surface density compared to that of the minimum mass model as the name of the model suggests. Although the disc mass is smaller, the gas accretion continues, and it brings enough amount of solid material for the satellite formation into the CJD. This continuous supply of gas is consistent with the recent view of the gas accretion process predicted by numerical simulations, but the continuous supply of solid material could be a problem (see Section \ref{supply}). In this model, it is assumed that the gas flows onto the CJD uniformly, and the outer edge of the gas inflow region is the centrifugal radius (i.e., $r_{\rm out}=r_{\rm c}$) similar to the minimum mass model. The value of the centrofugal radius is set as $r_{\rm c}=30~R_{\rm J}$, implicitly including the gravitational effect to the specific angular momentum unlike the minimum mass model. The disc is also assumed to be truncated at $r_{\rm trun}=150~R_{\rm J}$, considering the shock regions shown in the hydrodynamic simulations \citep[e.g.,][]{lub99,dan02}. This is also roughly consistent with an analytically derivation by \citet{mar11} that the tidal torques from the Sun removes the angular momentum of the gas at the outer region of the disc and make the disc truncated at $r_{\rm trun}\approx0.4~R_{\rm H}\approx300~R_{\rm J}$.

\citet{ali05} developed a similar accretion disk model being consistent with the formation process of Jupiter which the original model of \citet{can02} does not consider. \citet{war10} made a more detailed analytical model based on \citet{can02}. The satellite formation in the gas-starved disc model has been investigated well, and we will show that in Section \ref{formation}.

\subsection{Modern views of the CJD}
\label{modern}
The gas-starved disc model is the basis of the recent CJD/CPD models, but hydrodynamic simulations have revealed more detailed structures of the discs as the performance of the computers has been improved.

\subsubsection{Hydrodynamic simulations}
\label{hydro}
Hydrodynamic simulations\index{Hydrodynamic simulations of the CJD} have been carried out in many previous works. First, 2D isothermal simulations were calculated in some works \citep[e.g.,][]{lub99}, and then a lot of 3D isothermal simulations have been conducted \citep[e.g.,][]{kle01}. Self-gravitating and/or radiation hydrodynamic calculations resolving inside the Hill sphere have also been calculated \citep[e.g.,][]{ayl09a}. Ideal and non-ideal (including only Ohmic resistivity) magnetohydrodynamic (MHD) simulations\index{Magnetohydrodynamic simulations} have also been conducted by \citet{mac06} and \citet{gre13}, respectively (see also next section).

The gas around the core of Jupiter has an envelope-like structure when the mass of the planet is small, and the envelope becomes flatter as the planet grows heavier. The gas temperature is an important factor because the high temperature supports the gas structure against the gravity. The condition for the formation of the CJD has been investigated by some 3D radiation hydrodynamic simulations \citep[e.g.,][]{ayl09a}. For example, \citet{sch20} found that the flatness parameter of the envelope/disc starts to decrease (flatter) when the mass of proto-Jupiter reaches around $0.3~M_{\rm J}$, and also the envelope/disc becomes flatter as the opacity is lower. \citet{kra24} found that a necessary condition for the CPD formation is that the cooling time is at least one order of magnitude shorter than the orbital timescale.

\begin{figure}
\centering
\includegraphics[width=0.9\hsize]{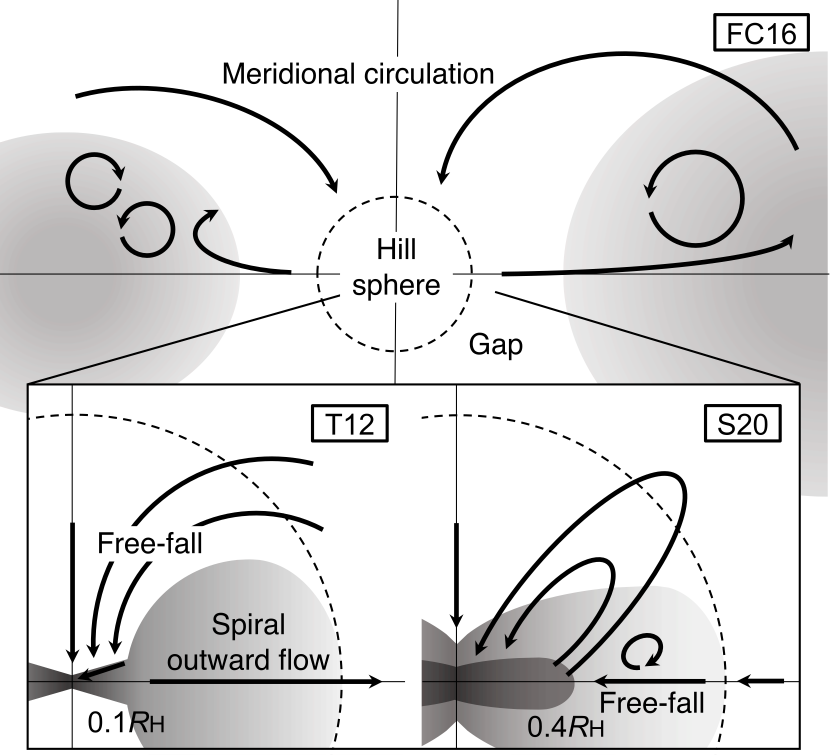}
\caption{Schematic picture of the azimuthally averaged gas structures around the CJD. The upper and lower panels represent the structures shown in the simulations by \citet{fun16} (FC16, isothermal), and by \citet{tan12} (T12, isothermal) and the low opacity case of \citet{sch20} (S20, radiation), respectively. The gray scales and black arrows roughly represent the gas density and the gas flows, respectively. \label{fig:gasflows}}
\end{figure}

After the formation of the CJD, the gas basically flows into the Hill sphere thorough the Lagrange points\index{Lagrange points}, ${\rm L_{1}}$ and ${\rm L_{2}}$, and is accreted into Jupiter thorough the CJD in spirals. However, the detailed structures are still controversial. We show a schematic picture of the gas structures in Figure \ref{fig:gasflows}. Large structures bridging over the gap named as ``meridional circulation\index{Meridional circulation}'', have been found in some hydrodynamic simulations; the gas flows from the edge of the gap to the Hill sphere thorough the high altitude \citep{mor14,fun16,szu22} (the upper part of Figure \ref{fig:gasflows}). In the case of isothermal 3D hydrodynamic simulations by \citet {tan12} (the left lower part of Figure \ref{fig:gasflows}), the gas cannot penetrate inside the Hill sphere (and the shock around its surface) from the mid-plane, but it goes up to the high altitude along the surface of the sphere and then flows down onto the surfaces of the CJD inside about $0.1~R_{\rm H}$ ($\approx100~R_{\rm J}$ with $M_{\rm cp}=M_{\rm J}$) with the free-fall speed. The fallen gas turns to Jupiter just after passing through the shock on the surfaces, because its specific momentum of the gas is smaller than that of the local Keplerian rotation. The effective outer edge of the gas inflow area is then $r_{\rm out}\approx0.03~R_{\rm H}$ ($\approx30~R_{\rm J}$), which is near to $r_{\rm out}$ assumed in \citet{can02}. The gas on the mid-plane outside about $0.1~R_{\rm H}$ flows outward in spirals and goes out from the Hill sphere through ${\rm L_{1}}$ and ${\rm L_{2}}$. It can be transported again to the CJD with the meridional circulation. Inside $0.03~R_{\rm H}$, the gas on the mid-plane rotates almost Keplerian speed. Although the gas flows have such complex structures, the radial profile of the gas surface density is smooth in the Hill sphere. On the other hand, \citet{sch19b,sch20} conducted 3D radiation hydrodynamic simulations with that the opacity level is 100 times lower than that of the interstellar medium (ISM) dust (i.e., dust-to-gas ratio is $10^{-4}$), which could be a realistic setting due to the filtering of solid particles by the gap (see Section \ref{filtering}). The simulations showed more complicated gas structures then that of \citet{tan12} (the right lower part of Figure \ref{fig:gasflows}). Most of the gas can penetrate inside the Hill sphere from the mid-plane around ${\rm L_{1}}$ and ${\rm L_{2}}$, and it flows inward with the free-fall speed until it collides with another shock around $0.4~R_{\rm H}$ from Jupiter. The gas can enter inside the shock from the mid-plane but is slowed down. The gas then goes up to the high altitude in spirals and falls down to the deeper inside region around $0.2~R_{\rm H}$. The gas density inside the shock is much higher than the outside. In the simulations with the settings of deeper gravitational potential of Jupiter or dust enriched gas, the temperature is higher, and the gas structures are similar to that of \citet{tan12} except for the outflows from the Hill sphere through ${\rm L_{1}}$ and ${\rm L_{2}}$ and consistent with other previous 3D radiation hydrodynamic simulations  \citep[e.g.,][]{lam19}. The inner high-density region inside $0.4~R_{\rm H}$ appears also in these cases. We note that this robust picture of the inner dense disc inside about $0.4~R_{\rm H}$ is consistent with the classical gas-starved disc model by \citet{can02} (see Section \ref{gas-starved}).

The large satellites form inside $0.1~R_{\rm H}$ in most of the formation scenarios (except for that of \citet{bat20} where satellites form in about $0.1-0.3~R_{\rm H}$), where the disc is more quiescent than the outside (see Section \ref{formation}). The direction of the gas flow on the mid-plane is important for the satellite formation, because the outward flow may avoid the loss of the solid materials to Jupiter, but it is still controversial (see Section \ref{satellitesimalformation}). Beside the two simulations by \citet{tan12} and \citet{sch20}, \citet{szu17b} found by 3D radiative hydrodynamic simulations that the gas flows inward inside $10 ~R_{\rm J}$, outward at $10-100~R_{\rm J}$, and again inward outside $100~R_{\rm J}$, but it depends on the viscosity (see also \citet{dra18b}). Non-ideal MHD simulations by \citet{gre13} showed that the gas flows inward and outward on the mid-plane inside and outside of $0.2~R_{\rm H}$, respectively, but the resolution is not enough for the calculations inside about $0.05~R_{\rm H}$.

In the inner regions, however, the temperature is higher than the ice sublimation point in any situations considered in the hydrodynamic simulations, which is not suitable for the formation of the icy moons (see also Section \ref{constraint}). Thus, the satellites should form later than the situations, when the amount of gas flowing into the Hill sphere has decreased significantly, and the CJD has been cooler. However, it is not clear that the gas structures obtained in the simulations are kept in such later situations. Moreover, the resolutions are not enough in most of the hydrodynamic simulations for investigating the detailed structures of the satellite formation region. Thus, hydrodynamic simulations with longer time ranges and higher resolutions are the important future works.

\subsubsection{Effects of magnetic fields}
\label{magneticfields}
Magnetic fields also play important roles in the formation and evolution of the CJD. Young Jupiter has a strong magnetic field\index{Magnetic field of Jupiter} and, in addition to that, the magnetic field lines vertically piercing the CSD should be drawn into the CJD with the gas accretion from the CSD to the CJD. Ionization of the CJD large enough for the magnetic field to affect the evolution of the disc is achieved by cosmic ray, stellar X-ray, and radionuclide \citep{fuj11,fuj14,tur14}. Global 3D MHD simulations\index{Magnetohydrodynamic simulations} of accreting planets have shown that the flow structure of the accreting gas is roughly the same as predicted by hydrodynamical simulations, but it also suggested the existence of the jet structures from around the center of the CJD like in the case of the star formation process \citep{mac06,gre13}. On the other hand, \citet{bat18} argued that the gas flows onto the top of Jupiter and flows down onto the mid-plane along with the magnetic field of the planet. The magnetic field of the young Jupiter should be stronger than the current magnetic field, and if the planet and the CJD are connected by the magnetic field, the angular momentum is transported from the planet to the disc, resulting in the slowdown of the spin of the planet and formation of the inner cavity \citep[e.g.,][]{tak96,chr09}. The inner cavity is also important for the satellite formation because it can stop their migration (see Section \ref{formation}).

The above works have not investigated the mechanism of the transport of the orbital angular momentum in the CJD. In the case of PPDs, the magneto-rotational instability (MRI) is one of the most important mechanisms \citep{bal91,bal98}. However, the MRI is not maintained in the CJD, because the magnetic diffusion is relatively quick due to the small size-scale of the CJD except for the very inner region where thermal ionization occurs \citep{fuj14,kei14,tur14,fuj17}. Therefore, the disc mass may be larger than that predicted by the gas-starved disc model, in which a normal strength of viscosity is assumed. We note that the CJD may evolve by the magnetic wind-driven accretion, which can decrease the gas surface density and make the disc laminar, as has been investigated in the context of PPDs \citep[e.g.,][]{bai15}. Indeed, \citet{shi23b} showed through 3D local non-ideal MHD simulations that magnetic wind-driven accretion can also occur in the CJD.

\subsubsection{Other effects}
\label{othereffects}
There are other factors that can modify the evolution of the CJD. Some 2D hydrodynamic simulations of the disc with radiative cooling show that two spiral shock arms\index{Spiral shock arms} induced by the tidal force from the Sun can transport the angular momentum and cause energy dissipation in the disc \citep{riv12,zhu16,che21}. The ability of the transport of the angular momentum is consistent with $\alpha\sim10^{-3}$ in the $\alpha$ disc model. However, this simulation is 2D, and the spiral structures may depend on the dimension of simulations, so future detailed 3D calculations is expected. Photoevaporation\index{Photoevaporation} plays an important role in the evolution of PPDs, and it can also affect the structures of the CJD \citep{ale06a,ale06b,mit11}. The gap around the disc is so deep that it was exposed to the interstellar far-ultraviolet (FUV) radiation, and the disc becomes truncated around the current orbit of Callisto \citep{obe20}. \citet{sch25} showed that self-shadowing of Jupiter has a significant effect on the CJD temperature, a drop of approximately $100~{\rm K}$ in the shadowed zone compared to the surrounding areas. Unfortunately, the comprehensive understanding of the CJD including all of the above effects has not been completed yet. There are still a lot of mysteries in the understanding of the disc.

\subsection{Simple 1D disc model and constraints form the icy moons}
\label{constraint}
We introduce a simple 1D disc model for the CJD to summarize the discussion. The diffusion equation for the gas surface density of the discs with continuous supplies of gas, $\Sigma_{\rm g}$, is given by \citep{sha73,fuj14},
\begin{equation}
\dfrac{\partial\Sigma_{\rm g}}{\partial t}=\dfrac{1}{r}\dfrac{\partial}{\partial r}\left[3r^{1/2}\dfrac{\partial}{\partial r}\left(r^{1/2}\nu\Sigma_{\rm g}\right)\right]+F_{\rm in},
\label{diffusion}
\end{equation}
where $r$, $\nu$, and $F_{\rm in}$ are the distance from the central plane (Jupiter), viscosity, and mass flux of the gas inflow. The viscosity is $\nu=\alpha c_{\rm s}H_{\rm g}$, where $\alpha$ is the strength of turbulence, $c_{\rm s}$ is the sound speed, and $H_{\rm g}=c_{\rm s}/\Omega_{\rm K}$ is the gas scale height where $\Omega_{\rm K}=\sqrt{GM_{\rm cp}/r^{3}}$ is the Kepler angular velocity. Here, based on the results of the hydrodynamic simulation by \citet{tan12}, we model $F_{\rm in}\propto r^{-1}$ for $r<r_{\rm out}$ and $F_{\rm in}=0$ for $r> r_{\rm out}$ (see the previous sections). Then, the stationary solution of the equation can be written as,
\begin{equation}
\Sigma_{\rm g}=\dfrac{\dot{M}_{\rm g}}{2\pi r_{\rm out}}\dfrac{r^{3/2}}{\nu}\left(-\dfrac{2}{9}r^{-1/2}+\dfrac{2}{3}r_{\rm out}r^{-3/2}\right),
\label{Sigmagf}
\end{equation}
where $\dot{M}_{\rm g}$ is the gas inflow rate onto the CJD, which is consistent with the (uniform) gas accretion rate inside the disc owing to the conservation of mass in the steady state. We note that if the gas flows only outside of $r<r_{\rm out}$ ($F_{\rm in}=0$ for $r<r_{\rm out}$), the gas surface density inside $r_{\rm out}$ can be,
\begin{equation}
\Sigma_{\rm g}=\dfrac{\dot{M}_{\rm g}}{3\pi\nu}=\dfrac{\dot{M}_{\rm g}\Omega_{\rm K}}{3\pi\alpha c_{\rm s}^{2}}.
\label{Sigmag}
\end{equation}
The sound speed depends on the disc temperature, $c_{\rm s}=\sqrt{k_{\rm B}T/m_{\rm g}}$, where $k_{\rm B}$, $T$, and $m_{\rm g}$ are the Boltzmann constant, disc temperature, and mean molecular mass of gas, respectively. The gas temperature on the mid-plane governed by the viscous heating is given by \citep{nak94},
\begin{equation}
T=\left(\dfrac{3GM_{\rm cp}\dot{M}_{\rm g}}{8\pi\sigma_{\rm SB}r^{3}}\right)^{1/4}g,
\label{T}
\end{equation}
where $\sigma_{\rm SB}$ is the Stefan-Boltzmann constant and,
\begin{equation}
g=\left(\dfrac{3}{8}\tau+\dfrac{1}{4.8\tau}\right)^{1/4}
\label{g-b}
\end{equation}
is a function of the Rosseland mean optical depth $\tau=\kappa\Sigma_{\rm g}$. In reality, the luminosity from Jupiter is another source of heating, which could be dominant in the outer region of the disc \citep{can02}. In principle, the opacity $\kappa$ depends on the size distribution of solid particles, but we assume $g=1$ for simplicity as a minimum estimate. The disc temperature is then determined by the gas accretion rate, $\dot{M}_{\rm g}$, and the gas surface density is regulated by $\dot{M}_{\rm g}$ and the strength of turbulence, $\alpha$.

\begin{figure}
\centering
\includegraphics[width=0.9\hsize]{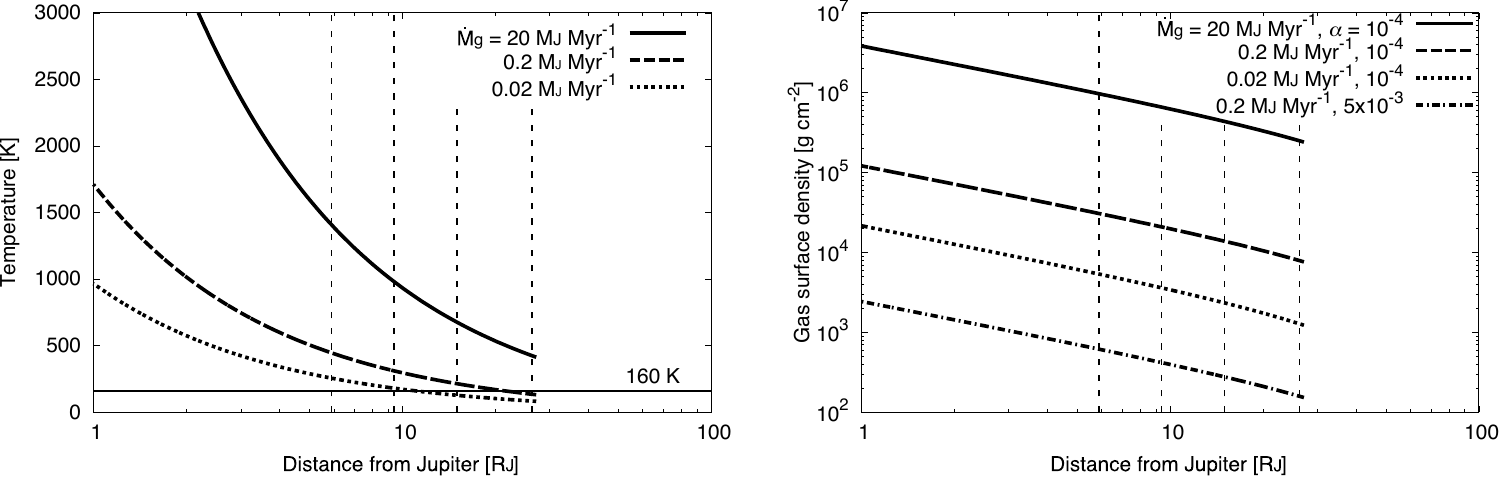}
\caption{{\bf Temperature and gas surface density of the CJD.} The curves in the left panel represent the mid-plane temperature of the CJD calculated by Eq. (\ref{T}) with the different gas accretion rates. The continuous, dashed, and dotted  curves represent the temperature with $\dot{M}_{\rm g}\sim20$, $0.2$, $0.02~M_{\rm J}~{\rm Myr}^{-1}$, respectively. The horizontal line is the sublimation temperature of water ice, $160~{\rm K}$. The right panel represents the gas surface density calculated by Eq. (\ref{Sigmagf}) with $r_{\rm out}=27~R_{\rm J}$. The continuous, dashed, and dotted curves are those with $\dot{M}_{\rm g}\sim20$, $0.2$, $0.02~M_{\rm J}~{\rm Myr}^{-1}$, respectively, and with $\alpha=10^{-4}$. The short dashed-dotted curve is that with $\dot{M}_{\rm g}=0.2~M_{\rm J}~{\rm Myr}^{-1}$ and $\alpha=5\times10^{-3}$, the same condition with \citet{can02}. The vertical dashed lines in both panels represent the current orbital positions of the Galilean moons. \label{fig:tempe}}
\end{figure}

\citet{can02} assumed that $\dot{M}_{\rm g}=0.2~M_{\rm J}~{\rm Myr}^{-1}$ and $\alpha=5\times10^{-3}$. In this case, from Eqs. (\ref{Sigmagf}) and (\ref{T}), the temperature and the gas surface density are the dashed curve in the left panel and the short dashed-dotted curve in the right panel of Figure \ref{fig:tempe}, respectively. \citet{shi17} assumed that $\dot{M}_{\rm g}=0.02~M_{\rm J}~{\rm Myr}^{-1}$ and $\alpha=10^{-4}$, which are the dashed-dotted curve in the left panel and the dashed-dotted curve in the right panel of the figure, respectively. In this case, the snow line is around the current orbit of Europa, which is suitable in order to explain the ice mass fractions of the satellites (see Section \ref{icefractions}).

On the other hand, 3D hydrodynamic simulations show that if the gas surface density of the CSD is typical, the gas accretion rate is the order of $\dot{M}_{\rm g}\sim20~M_{\rm J}~{\rm Myr}^{-1}$ even if the gap structure is open \citep{lam19}. In this case, the disc temperature around the current orbits of the Galilean moons is much higher than the sublimation temperature of water ice, about $160~{\rm K}$ (see the continuous curve in the left panel of Figure \ref{fig:tempe}). This is inconsistent with the existence of the icy moons if their solid materials formed in-situ. Moreover, such a high accretion rate, by definition, cannot remain for more than $\sim0.01~{\rm Myr}$ except if a very strong fraction of this mass is recycled back to the CSD. Therefore, the satellite formation should have occurred in the final phase of the evolution of the CSD, when the gas surface density of the CSD has decreased, and so the gas accretion rate has become lower. On the other hand, there are some research works which proposed the satellite(simal) formation at the outer regions of the CJD where the temperature is low enough even with the high gas accretion rate \citep{cil18,dra18b,bat20,cil21}.

\section{Supply of solid materials to the birthplace of Ganymede}
\label{supply}
In the previous section, we have explained the current understanding of the structure of the CJD, where the large satellites including Ganymede formed. Before we explain how Ganymede formed in the CJD, we have to discuss an important factor in the satellite formation, the supply of solid material to the disc.

\subsection{Filtering of pebbles by gap structure}
\label{filtering}
In the classical satellite formation models, it is assumed that enough solid material is supplied to the CJD. In the case of the gas-starved disc model, the dust-to-gas mass ratio of the gas inflow is a parameter regulating the total mass of the large satellites of the systems \citep{can06}. However, it has been shown that the amount of dust supply to the CJD is not a free parameter but is very limited.

In order to keep the disc temperature cold enough for the formation of the icy moons, the mass flux of the gas accretion from the CSD to CJD needs to be small, which is only possible when the gas surface density of the CSD has decreased enough (see Section \ref{constraint}). At such a late stage of the CSD evolution, the gap structure has also already formed around the CJD. However, the gap structure filters the pebble size particles and most of the solid materials cannot enter inside the gap and not be transferred to the CSD \citep[e.g.,][]{paa04,zhu12,hom20}\index{Filtering of pebbles}. Dust particles in the CSD can grow to pebbles (about centimeter to meter size) by collisional growth, but they drift inward at the size. The negative gas pressure gradient of the CSD supports the gas disc and so the gas rotates around the star with sub-Kepler speed, which becomes the head wind against the particles trying to rotate with Kepler speed. As a result, the particles lose their angular momentum and drift inward. The efficiency of losing angular momentum gets higher when the particles have the size of pebbles. However, at the outer edge of the gap structure, the gas pressure gradient is positive, and pebbles drift outward. Therefore, the pebbles do not enter inside the gap and pile up at the gas pressure maximum unless their diffusion is so strong and push them against the drift mechanism.

An alternative way to transport the materials to the CJD is changing the size of the particles to be so large that the particles do not suffer the effects of the gas pressure, or to be so small that the particles can enter the gap coupled with the gas. Some recent works argued that the pebbles piling up at the gas pressure bump\index{Gas pressure bumps} can form planetesimals by the streaming instability\index{Streaming instability} and so on \citep[e.g.,][]{lyr09,ayl12a,cha13,tak16,sta19,eri20,shi20,shi23a}. Moreover, a lot of small particles can also form as the fragments out of the collisions of the planetesimals \citep{kob12}. Therefore, these small dust particles and large planetesimals may be supplied to the CJD. We discuss the above two ways of the transport in the next two subsections.

\begin{figure}
\centering
\includegraphics[width=0.9\hsize]{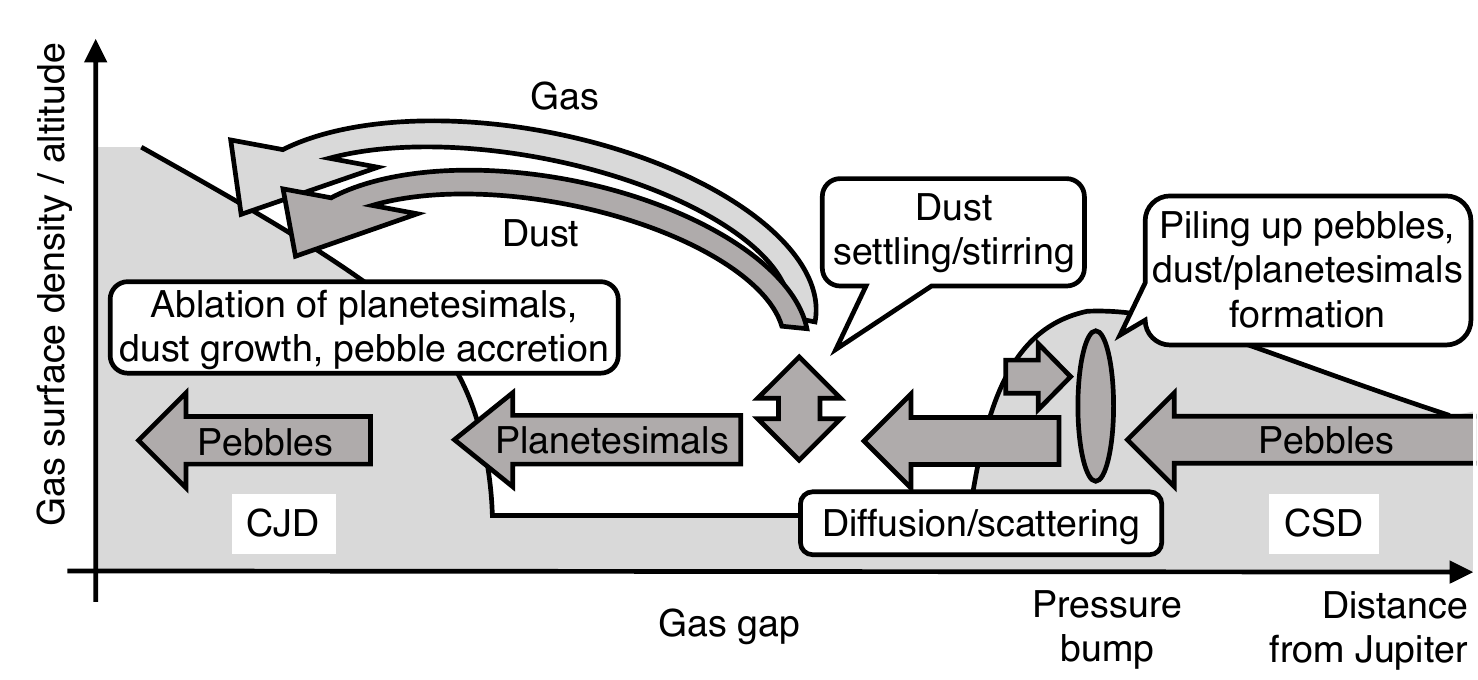}
\caption{{\bf Schematic picture of transport of materials to the CJD.} The horizontal axis roughly represents the distance from Jupiter. The background black solid curves and the gray shaded regions represent the gas surface density. The solid material of the moons originates at the outer region of the CSD and drift inward as pebbles. The drifting pebbles then pile up at the gas pressure bump created by Jupiter. Planetesimals form from the gathered pebbles by the streaming instability, and their fragmentation also forms small dust particles. Planetesimals scattered by large bodies (such as Saturn) get high eccentricity and penetrate the gap. Small dust particles can also penetrate the gap by their diffusion if it is strong, and they are supplied onto the CJD with the gas inflow from the high altitude, which also depends on their vertical settling and stirring inside the gap. The supplied dust particles grow to pebbles and drift inward as pebbles inside the CJD. The captured planetesimals are also ablated and produce dust particles. They also grow by accreting the drifting pebbles. \label{fig:transport}}
\end{figure}

\subsection{Supply of dust particles coupled with inflowing gas}
The idea of transporting solid materials to the CJD as small dust particles has been assumed implicitly in the classical works \citep{can02}. In most of the formation scenarios, the solid-to-gas mass flux ratio supplied to the CJD is an important parameter which governs the total mass of satellites formed by satellitesimal accretion. \citet{hom20} and \citet{mae24} found that most of the dust particles around the CJD can flows onto the disc with gas if there is enough strength of turbulence to blow up the dust particles to the high altitude where the gas flows toward the CJD as predicted in \citet{tan12}.

However, when the gap structure exists, in other words, when the large satellites form in the CJD, most of the particles cannot access to the root of the gas inflow which is inside the gap. Only small particles can penetrate inside the gap thanks to the diffusion and be supplied to the CJD, which should make the dust-to-gas ratio much lower than that expected in the previous satellite formation works \citep{paa06,paa07,zhu12}. The condition for the penetration can be roughly explained as follows. The radial velocity of particle in gas is given by,
\begin{equation}
v_{\rm r}=v_{\rm drift}+v_{\rm diff},
\label{vr}
\end{equation}
where $v_{\rm drift}$ and $v_{\rm diff}$ are the drift velocity\index{Drift of pebbles} and the diffusion velocity, respectively. The drift velocity is \citep{ada76,wei77},
\begin{equation}
v_{\rm drift}=-2\dfrac{{\rm St}}{{\rm St}^{2}+1}\eta v_{\rm K},
\label{vdrift}
\end{equation}
where ${\rm St}\equiv\tau_{\rm stop}/\Omega_{\rm K}$ is the Stokes number\index{Stokes number} of the particle, where $\tau_{\rm stop}$ is the stopping time, and $v_{\rm K}=r\Omega_{\rm K}$ is the Kepler velocity. The ratio of the pressure gradient force to the gravity of the central planet is
\begin{equation}
\eta=-\dfrac{1}{2}\left(\dfrac{H_{\rm g}}{r}\right)^{2}\dfrac{\partial \ln{\rho_{\rm g}c_{\rm s}^{2}}}{\partial \ln{r}},
\label{eta}
\end{equation}
where $\rho_{\rm g}=\Sigma_{\rm g}/(\sqrt{2\pi}H_{\rm g})$ is the gas density at the mid-plane. At the outer edge of the gap, $\eta$ is negative. The diffusion velocity is
\begin{equation}
v_{\rm diff}=-\dfrac{D}{r}\dfrac{d\ln{(\Sigma_{\rm d}/\Sigma_{\rm g})}}{d\ln{r}}.
\label{vdiff}
\end{equation}
The turbulent diffusivity of the particles is $D=\nu/{\rm Sc}$, where the Schmidt number is ${\rm Sc}=(1+{\rm St}^{2})^{2}/(1+4{\rm St}^{2})$ \citep{sha73,you07}. When $|v_{\rm drift}/v_{\rm diff}|<1$, the particles can penetrate the gap. From the above equations,
\begin{equation}
\left|\dfrac{v_{\rm drift}}{v_{\rm diff}}\right|\sim\dfrac{\rm St}{\alpha}\left(\dfrac{d\ln{\Sigma_{\rm g}}}{d\ln{r}}\right)\left(\dfrac{d\ln{(\Sigma_{\rm d}/\Sigma_{\rm g})}}{d\ln{r}}\right)^{-1},
\label{vbalance}
\end{equation}
so the rough condition for the penetration of the particles is ${\rm St}/\alpha\ll1$ \citep{zhu12}. On the other hand, the other terms depend on the gas gap structure which is determined by the mass of the planet, the aspect ratio of the CSD, and so on. When ${\rm St}$ and $\alpha$ are given, the minimum planet mass which the pebbles cannot penetrate the gap with (i.e., when $v_{\rm drift}/v_{\rm diff}|=1$) is called ``pebble isolation mass\index{Pebble isolation mass}'' \citep[e.g.,][]{lam14,ata18}. If planets glow larger than their pebble isolation mass, they cannot accrete pebbles anymore (see also Section \ref{pebaccretion}).

We note that a recent work, \citet{szu22}, carried out 3D dust+gas radiative hydrodynamic simulations and found that, even with the gap structure, the meridional circulation drives a strong vertical flow even for mm-sized grains and can deliver them to the CJD with the gas from the high altitude (see Figure \ref{fig:gasflows}). In addition, the gas easily flows out from the Hill sphere, but the dust is easier trapped in the planet region. As a result, the accretion rates onto the vicinity (inside $0.5~R_{\rm H}$) of Jupiter for the dust can be $\sim0.01~M_{\rm J}~{\rm Myr}$, and that for the gas is $\sim0.1~M_{\rm J}~{\rm Myr}$. We also note that the observation of the CPD around PDS~70~c shows that there are a lot of dust particles inside the CPD \citep{ise19,ben21,por22,cas22,fas25,dom25}. \citet{chr24} also found a stream like structure of dust emission bridging the outer ring of PDS~70 system and the vicinity of PDS~70~c by the James Webb Space Telescope (JWST). However, in the case of a heavy planet ($\sim10~M_{\rm J}$), its PPD may turn eccentric, and the gap edge could approach the CPD and the planet, resulting in an increase of gas accretion rate \citep{kle06}, which may make it possible to supply a lot of dust with the gas. Therefore, it is difficult to define whether this observation of PDS~70~c supports the argument that plenty of dust can be transported to the CJD or not.

\subsection{Capture of planetesimals by the CJD}
\label{capture}
The idea of the capture of planetesimals\index{Capture of planetesimals} by the gas drag from CPDs has been investigated a lot \citep{fuj13,tan14,dan15,sue16a,sue16b,sue17,ron18,ron20}. In the case of calculations of \citet{ron20}, about $23\%$ of the planetesimals in the feeding zone of Jupiter can be captured by the CJD. \citet{sue17} calculated the capture and subsequently orbital evolution of planetesimals and found that the initial radial distribution of satellitesimals assumed in the satellitesimal accretion scenarios (see Section \ref{sataccretion}) could be roughly reproduced by the prograde captured planetesimals. Their surface number density is significantly enhanced at the current location of the Galilean moons. There is also another type of captured planetesimals which take ``long-lived'' prograde orbits \citep{sue16a}, which are captured at farther orbits from Jupiter and could be the seeds of the Galilean moons (see \citet{shi19} and Section \ref{pebaccretion}). When planetesimals get close to Jupiter, their surface temperature increases, and the ablation\index{Ablation} (mass loss of the surface material by sublimation and by melting and evaporation) is most vigorous. Some works argued that the ablation (and break-up) of the planetesimals made the CJD dust-laden \citep{dan15,ron18}. Especially, in the pebble accretion scenario by \citet{ron20}, the dust particles released by the ablation of the captured planetesimals drift inward as pebbles and are accreted onto the large proto-satellites which are the planetesimals survived the ablation (see Section \ref{pebaccretion}).

However, it might be difficult to supply planetesimals to the CJD continuously. In the case of \citet{ron20}, the planetesimals inside the feeding zone have been cleaned up in $\sim10^{4}~{\rm yr}$. Moreover, the gas pressure gradient created by Jupiter and the gravitational interaction with the growing planet itself should push the planetesimals out from Jupiter's feeding zone, rendering the planetesimal capture rate very low \citep{hay77,fuj13}. On the other hand, the pebbles piled up at the gas pressure bump form planetesimals continuously thanks to the continuous supply of pebbles drifting from the outer region of the CSD to the bump \citep{shi20}. Massive embryos then form by the runaway growth of the planetesimals and can excite the planetesimals orbits, resulting in the penetration of some planetesimals inside the gap \citep{kob12}. Moreover, the outer planet, Saturn, can destabilize the planetesimals and deliver them to the feeding zone, which is an enough amount for the formation of the Galilean moons \citep{ron18,ron20}.

\section{Formation of Ganymede and the Galilean moons}
\label{formation}
In the previous sections, we discussed the possibilities to transport solid materials from the CSD to the circum-Jovian disc. In this section, we show how Ganymede and the other large moons formed from the materials in the disc. First, we explain ``satellitesimal accretion'' scenario, which have been considered as an orthodox way to form the large moons. Then, we show recent alternative ``pebble accretion'' scenarios, which may avoid the problems of the satellitesimal accretion scenarios.

\subsection{Satellitesimal accretion scenarios}
\label{sataccretion}
Satellitesimals\index{Satellitesimal}, the ``satellite version'' of planetesimals have been considered as the building blocks of the large moons. First, the formation process from satellitesimals to the large satellites were investigated. The basic idea of the satellite accretion scenario\index{Satellitesimal accretion} is as follows. There is always a sufficient number of satellitesimals inside the CJD thanks to the continuous supply of dust particles, resulting in steady satellite formation \citep{can06,sas10,ogi12}. However, as we discuss in the previous sections, the supply of dust particles is always not guaranteed to be large enough. Moreover, it has been shown that the satellitesimal formation from dust particles in CPDs have many issues like the planetesimal formation in PPDs, which was not discussed in the classical satellitesimal accretion research works. In this subsection, we first explain the issues and proposed solutions of the formation of satellitesimal. Then, we show the satellite formation processes starting from the satellitesimals (i.e., satellitesimal accretion and migration) based on classical studies. We also show the schematic pictures of the satellitesimal accretion scenarios in Figure \ref{fig:satelltesimals}, including the classical minimum mass model (see Section \ref{MMSD}) at the left upper part of the figure.

\subsubsection{Satellitesimal formation in the CJD}
\label{satellitesimalformation}
In most of the satellite formation\index{Satellitesimal formation} works based on the satellitesimal accretion scenarios, the existence of the satellitesimals has been assumed \citep{can06,sas10,ogi12}. However, it has been known that dust particles grow to pebbles but drift\index{Drift of pebbles} towards the central stars, and it is difficult to form planetesimals from the pebbles in PPDs unless in some special conditions (around snow lines, at gas pressure bumps, etc.). The gas pressure supports the gas and make it rotate slower than the Kepler speed. The pebbles trying to rotate with Kepler speed then receive head wind from the gas and lose their angular momentum, resulting in their radial drift. This mechanism also works for the particles in the CJD. \citet{shi17} modeled 1D simple CPDs and calculated the simultaneous growth and drift of the dust particles which have single peak mass at each distance from the planet. First, from the conservation of the mass,
\begin{equation}
\dot{M_{\rm d}}=-2\pi rv_{r}\Sigma_{\rm d},
\label{continuous}
\end{equation}
where $v_{\rm r}\approx v_{\rm drift}$, and $v_{\rm diff}$ can be ignored (see Eq. (\ref{vr})). The collisional growth\index{Collisional growth} under the steady condition is \citep{sat16,shi17},
\begin{equation}
v_{\rm r}\dfrac{d m_{\rm d}}{d r}=4\pi R_{\rm d}^{2}\rho_{d}\Delta v_{\rm dd}\epsilon_{\rm grow}
\label{growth}
\end{equation}
where $R_{\rm d}$ is the dust radius. The dust-dust relative velocity (i.e., collision velocity) is
\begin{equation}
\Delta v_{\rm dd}\approx\sqrt{(v_{\rm r}/2)^{2}+v_{\rm t}^{2}},
\label{vdd}
\end{equation}
where the turbulent-driven velocity $v_{\rm t}=\sqrt{3\alpha{\rm St}}c_{\rm s}$ \citep{orm07,sat16}. The particle density on the mid-plane is \citep{you07},
\begin{equation}
\rho_{d}=\dfrac{\Sigma_{\rm d}}{2\sqrt\pi H_{\rm d}}=\dfrac{\Sigma_{\rm d}}{2\sqrt\pi H_{\rm g}}\left(1+\dfrac{\rm St}{\alpha}\dfrac{1+2{\rm St}}{1+{\rm St}}\right)^{1/2},
\label{rhod}
\end{equation}
where $H_{\rm d}$ is the dust scale height. The Stokes number in the CJD shifts from the Stokes to Newton regimes as the dust grow large,
\begin{equation}
{\rm St}=\dfrac{8}{3C_{\rm D}}\dfrac{\rho_{\rm int}R_{\rm d}}{\rho_{\rm g}\Delta v_{\rm dg}}\Omega_{\rm K},
\label{stokes_newton}
\end{equation}
where $C_{\rm D}\approx0.5$ (in the Newton regime), $\rho_{\rm int}$, and $\Delta v_{\rm dg}$ are a dimensionless coefficient, the internal density of the particles, and the relative velocity between the particles and gas. The fragmentation\index{Fragmentation of pebbles} occurs when the collision speed is too high. The sticking efficiency is then \citep{oku16},
\begin{equation}
\epsilon_{\rm grow}=\min\left\{1, -\dfrac{\ln{(\Delta v_{\rm dd}/v_{\rm cr})}}{\ln{5}}\right\},
\label{egrow}
\end{equation}
where the critical velocity, $v_{\rm cr}$, for the collision of the rocky and icy particles are about $v_{\rm cr}=5$ and $50~{\rm m~s^{-1}}$, respectively \citep{wad09,wad13}.

\begin{figure}
\centering
\includegraphics[width=0.9\hsize]{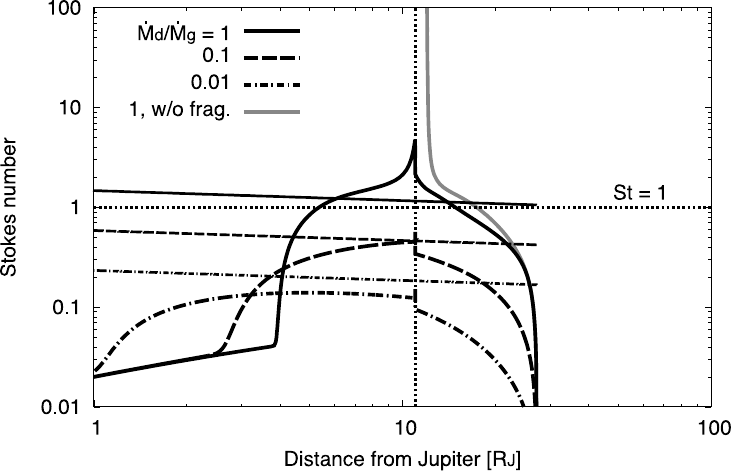}
\caption{{\bf Evolution of dust particles in the CJD.} The black continuous, dashed, and dashed-dotted curves represent the Stokes number of the dust when $\dot{M}_{\rm d}/\dot{M}_{\rm g}=1$, $0.1$, and $0.01$, respectively. The gray curve is the case without the fragmentation (i.e., $\epsilon_{\rm grow}$ is fixed as unity). The thin lines represent the Stokes number estimated by Eq. (\ref{St}). The horizontal and vertical dotted lines represent ${\rm St}=1$ and the position of the snowline, respectively. \label{fig:evol}}
\end{figure}

Figure \ref{fig:evol} shows the evolution of dust particles in the turbulent CJD calculated by the above equations. The gas surface density is calculated by Eq. (\ref{Sigmagf}) with $r_{\rm out}=27~R_{\rm J}$, $\dot{M}_{\rm g}=0.02~M_{\rm J}~{\rm Myr}^{-1}$, and $\alpha=10^{-4}$. We assume that all of the dust particles are supplied at $r_{\rm out}$ to carry out the single size calculation. The particles can grow to large icy satellitesimals only when the fragmentation is not considered (i.e., $\epsilon_{\rm grow}$ is fixed as unity) and $x\geq1$, where $x\equiv\dot{M}_{\rm d}/\dot{M}_{\rm g}$ is the dust-to-gas mass ratio of the accretion inflow, and this condition is virtually unattainable (see also \citet{shi17}). Furthermore, if the fragmentation is considered, the particles cannot grow to planetesimals but drift inwards even when $x=1$. The sizes of the particles decrease as they move inward especially inside the snowline because the fragmentation prevents their growth due to the slower critical fragmentation velocity of rocky particles than that of icy particles. The figure also shows that the particles can glow large as the dust-to-gas mass ratio of the inflow is large. When the fragmentation is inefficient ($\epsilon_{\rm grow}=1$), the Stokes number of the drifting pebbles can be estimated by taking a limit of $r\to0$ of Eq. (\ref{growth}),
\begin{equation}
{\rm St}\approx1.2\left(\dfrac{x}{1}\right)^{2/5}\left(\dfrac{\alpha}{10^{-4}}\right)^{1/5}\left(\dfrac{T}{160~{\rm K}}\right)^{-2/5}\left(\dfrac{M_{\rm cp}}{1~M_{\rm J}}\right)^{2/5}\left(\dfrac{r}{10~R_{\rm J}}\right)^{-2/5},
\label{St}
\end{equation}
where $M_{\rm cp}$ is the mass of the central planet. The drift speed becomes the fastest when the Stokes number of the pebbles is unity (see Eq. (\ref{vdrift})), so if the Stokes number becomes larger than unity, the pebbles start to pile up at that point, and satellitesimals can form there, which is similar to what happens in the planetesimal formation in PPDs. Therefore, Eq (\ref{St}) shows that the condition for the satellitesimal formation is $x\geq1$ (if the fragmentation is ignored), and as $x$ is large, the particles can glow large, which are consistent with the results of the calculations. The streaming instability\index{Streaming instability}, which is driven by the difference between the velocities of the dust and gas, is one of the most effective ways to avoid the radial drift and form planetesimals in PPDs \citep{you05}, but it does not work in the CJD because the dust-to-gas density ratio on the mid-plane is much smaller than the critical value for the instability, unity \citep{joh07,dra14}. Moreover, the internal density of the particles does not affect the results although the fluffy particles may be able to solve the radial drift issue in PPDs \citep{oku12}.

If the gas flows outward on the mid-plane in spirals, it may stop the inward drift of pebbles. Although the disc model used in the above discussion is a smooth accretion disc, a previous 3D hydrodynamic simulation showed that the gas on mid-plane at $r\approx10-100~R_{\rm J}$ should flow outward \citet{szu17b}. In that case, the dust particles (pebbles) pile up around $85~R_{\rm J}$ and the dust-to-gas ratio becomes high enough to trigger the streaming instability \citep{dra18b}. \citet{bat20} argued that satellitesimals form by the gravitational instability at the outer region of the CJD, $r\approx100-300~R_{\rm J}$, where the gas on the mid-plane flows outward according to some previous works of the CJD \citep[e.g.,][]{tan12,bat18}. The radial drift of the small particles is stopped by the outward flow and the dust-to-gas ratio increases enough for the occurrence of the gravitational instability \citep{bat20}. Based on those discussion, recently some scenarios of satellitesimal accretion at the outer region of the CJD has been proposed (see Section \ref{outerregion}). Also, the efficiency of the collisional growth of the particles is higher if the disc is laminar (i.e., turbulence does not occur), for example, with the magnetic wind-driven accretion \citep{shi23b}. The increase of the efficiency also occurs in the cases of the collisional growth in PPDs \citep{dra18a}.

\subsubsection{Satellitesimal accretion at the inner region of the CJD}
\label{satellitesimalaccretion}
The formation process from satellitesimals to satellites is also similar to that from planetesimals to planets. The satellites cannot grow infinitely large because the Type I migration\index{Migration of satellites} speed of satellites becomes faster as they become larger \citep[e.g.,][]{tan02}, which governs the final mass of the large moons if there is no structure which can stop the migration \citep{can06}. Here we show the estimate of the mass of the moons by \citet{can06}. Thanks to the continuous dust supply, the formation and migration of the satellites repeat until the supply of dust and gas stops. The current satellites are the final generation of the satellites repeating formation and migration. The growth timescale of the (proto-)satellite is,
\begin{equation}
\tau_{\rm acc}=\dfrac{M_{\rm s}}{2\pi r\Delta rxF_{\rm in}}
\label{tacc}
\end{equation}
where $M_{\rm s}$, $F_{\rm in}$, and $2\pi r\Delta r$ are the (proto-)satellite mass, the gas inflow flux per area, and the annular disc area over which the satellite accumulates the solid material, respectively. We assume that $F_{\rm in}$ and $x$ are both unity and constant. We also note that this equation is given under the assumption that the supplied dust particles are immediately accreted to the satellites, which neglects the inward drift of particles we discussed in the previous section. The annulus width is given by $\Delta r/r\approx2e$, where $e$ is the characteristic maximum eccentricity of the satellite. The eccentricity is given by $e\approx(H_{\rm g}/r)(M_{\rm s}/4\pi rH_{\rm g}\Sigma_{\rm g})^{1/5}$, which comes from the balance between the eccentricity damping by density waves and the excitation by gravitational scatterings with other satellitesimals \citep{war93}. The migration timescale is,
\begin{equation}
\tau_{\rm mig}=\dfrac{1}{C_{a}}\left(\dfrac{M_{\rm cp}}{M_{\rm s}}\right)\left(\dfrac{M_{\rm cp}}{r^{2}\Sigma_{\rm g}}\right)\left(\dfrac{H_{\rm g}}{r}\right)^{2}
\label{tmig}
\end{equation}
where $C_{a}$ is the constant of Type I migration order of unity \citep{tan02}. From the balance between the two timescales, the critical mass of satellites $m_{\rm crit}$ can be given by
\begin{equation}
\dfrac{m_{\rm crit}}{M_{\rm cp}}\approx5.6\times10^{-5}\chi\left(\dfrac{3.5}{C_{a}}\right)^{5/9}\left(\dfrac{H_{\rm g}/r}{0.1}\right)^{26/9}\left(\dfrac{r/r_{\rm out}}{0.5}\right)^{10/9}\left(\dfrac{\alpha}{3\times10^{-3}}\right)^{2/3}\left(\dfrac{x}{0.01}\right)^{2/3},
\label{mcrit}
\end{equation}
where $\chi$ is order of unity with very weak dependence of the gas accretion rate. Then, the total mass of satellites, $M_{\rm T}$, is,
\begin{equation}
\begin{split}
\dfrac{M_{\rm T}}{M_{\rm cp}}&\approx\int^{r_{\rm out}}_{R_{\rm p}}\dfrac{m_{\rm crit}/M_{\rm cp}}{\Delta r}{\rm d}r \\
&\sim2.5\times10^{-4}\dfrac{1}{\chi}\left(\dfrac{3.5}{C_{a}}\right)^{4/9}\left(\dfrac{H_{\rm g}/r}{0.1}\right)^{10/9}\left(\dfrac{\alpha}{3\times10^{-3}}\right)^{1/3}\left(\dfrac{x}{0.01}\right)^{1/3}.
\label{MT}
\end{split}
\end{equation}
Therefore, the key parameters are $\alpha$ and $x$, in other words, the efficiency of transport of gas angular momentum and the dust-to-gas mass ratio of the gas inflow onto the CJD, which is similar to the situation of satellitesimal formation. At the same time, the dependence of the parameters is so weak in this scenario that the ratio between the total mass of the Galilean satellites and Jupiter, $10^{-4}$, can be explained naturally (provided enough amount of dust is supplied and not lost from the CJD). This ratio is common in the satellite systems in our solar system, the systems of Saturn and Uranus. The right upper part of Figure \ref{fig:satelltesimals} illustrates this scenario. We note that there is another scenario that the captured planetesimals are used as the alternatives of the satellitesimals \citep{sue17} (see Section \ref{capture}), which is shown in the right middle part of the figure.

Some research works invoked an inner cavity of the CJD, a truncation of the gas disc. It is formed by the magnetic coupling between the disc and Jupiter and could stop the migration of the satellites \citep{ogi10,liu17}. The magnetic coupling can also explain the reason why the spin rate of Jupiter became sufficiently slower than the break-up spin rate \citep{tak96,bat18}. If the inner cavity exists, the inward migration of the inner most satellite stops at the edge of the disc. The outer satellites approach inwardly from the outer region, and if the migration speed of the approaching satellites is slow enough, the outer satellites are captured into mean-motion resonances\index{Mean motion resonances} (MMRs) with the next inner satellite one by one. \citet{sas10} and \citet{ogi12} performed semi-analytical Monte Carlo simulations and N-body simulations, respectively. They found that four to five similar-mass satellites are likely to remain in the Jovian system trapped in MMRs. This result is very similar to the current orbital characteristics of the Galilean moons; Io-Europa and Europa-Ganymede are in the 2:1 mean motion resonances, but Ganymede-Callisto are not in any resonances (see also Section \ref{orbit}). We note that the condition for maintaining the resonance chain of satellites is not determined by the migration timescale but by the total mass of the CJD \citep{sas10}, or by the total number of the satellites in the resonance chain \citep{ogi12}. When the trapped satellites have the critical values, the inner most satellite is pushed into the central planet by the outer satellites. The left middle part of Figure \ref{fig:satelltesimals} represents this scenario.

\subsubsection{Satellitesimal accretion at the outer region of the CJD}
\label{outerregion}
\begin{figure}
\centering
\includegraphics[width=0.9\hsize]{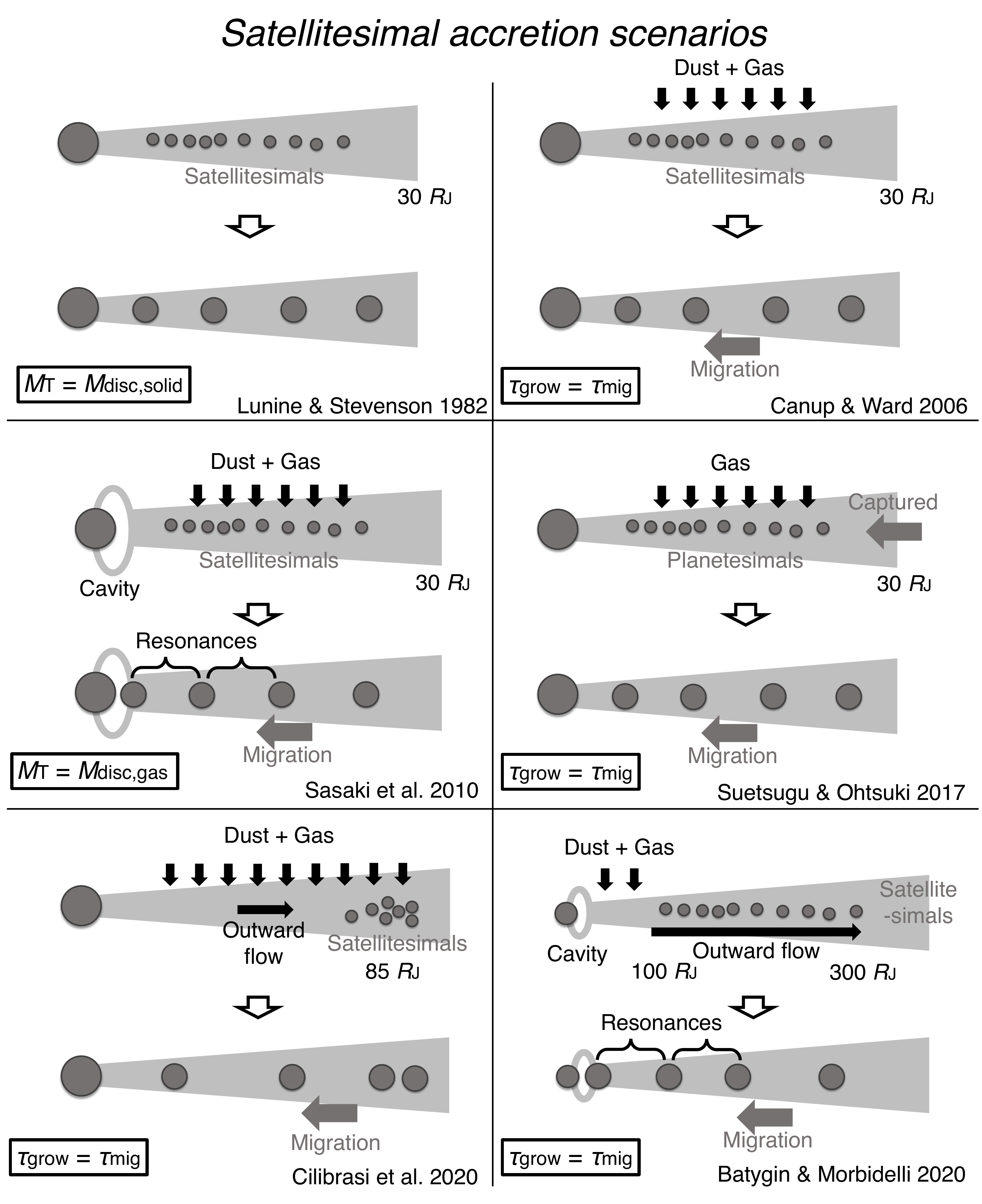}
\caption{{\bf Schematic pictures of the satellitesimal accretion scenarios.} The black boxes represent how the final mass of the moons is determined in each scenario. The total mass of the moons, the total mass of solid material in the CJD, and the gas disc mass are $M_{\rm T}$, $M_{\rm disc,solid}$, and $M_{\rm disc,gas}$, respectively. The timescale of growth and migration of the satellites are $\tau_{\rm grow}$ and $\tau_{\rm mig}$, respectively. The representative reference papers are also shown. \label{fig:satelltesimals}}
\end{figure}

In the above scenarios, satellitesimal accretion occurs inside about $30~R_{\rm J}$, in other words, around the current orbits of the Galilean moons. However, some recent works argued that the places of the satellitesimal accretion could be more outside regions of the CJD. \citet{cil18} investigated a scenario that the proto-satellites form around $85~R_{\rm J}$, where the dust particles pile up owing to that the direction of gas flow on the mid-plane switches from inward to outward there \citep{dra18b}. Thanks to the concentration of the solid materials, a lot of satellites form quickly ($\sim10^{4}~{\rm yr}$) there, but they then migrate inward by Type I migration as they grow larger. The formation and migration of satellites repeatedly occur, and their final generation becomes the Galilean moons. The final mass of the moons is determined by the balance between the growth timescale and the Type I migration timescale, like the classical scenario by \citet{can06}. In order to trigger the streaming instability, \citet{cil18} assumed high dust-to-gas accretion rates onto the CJD (0.03-0.5), resulting in a  peak of the distribution of the satellite-to-planet mass ratio locates between $10^{-4}$ and $10^{-3}$, which is heavier than the total mass of the Galilean satellites. However, as we discussed in the previous sections, getting high value of the dust-to-gas ratio is difficult. The left bottom part of Figure \ref{fig:satelltesimals} represents this scenario.

Much more outer regions could also be the birthplaces of satellitesimals and satellites. \citet{bat20} argued that $\sim100~{\rm km}$ sized satellitesimals are able to form at $r\approx0.1-0.3~R_{\rm H}$ by gravitational instability because dust particles can avoid their radial drift owing to the outward gas flow on the mid-plane, and the dust-to-gas ratio can be high. Satellite embryos emerge and grow to larger sizes by accreting the satellitesimals. The embryos also migrate inward, and their growth by satellitesimal accretion stops when they go outside the satellitesimal forming region. The inward migration stops at the inner edge of the disc formed by the magnetic field of Jupiter, which is similar to some of the previous works. The inner three moons, Io, Europa, and Ganymede are captured into the 2:1 MMRs one by one, and Callisto forms after the gas disappears from the CJD, which is the reason why the satellite is not in any resonances in this scenario. Moreover, Callisto accretes satellitesimals slowly owing to the excitation of their orbital distribution (see also Section \ref{constraints}). The right bottom part of Figure \ref{fig:satelltesimals} represents this scenario.

\subsection{Pebble accretion scenarios}
\label{pebaccretion}
In the field of the planet formation, ``pebble accretion\index{Pebble accretion}'' scenario has been investigated in many studies \citep[e.g.,][]{orm10,lam12}. As we discussed in Section \ref{satellitesimalformation}, dust particles in PPDs and in CPDs drift inward owing to the headwind they receive from the gas discs rotating with sub-Kepler speed, lose their angular momentum, and drift inward as centimeter or meter sized particles called ``pebble\index{Pebbles}'' before they grow larger to planetesimals/satellitesimals. We introduce here satellite formation scenarios which use these pebbles as the building blocks of the moons.

\begin{figure}
\centering
\includegraphics[width=0.9\hsize]{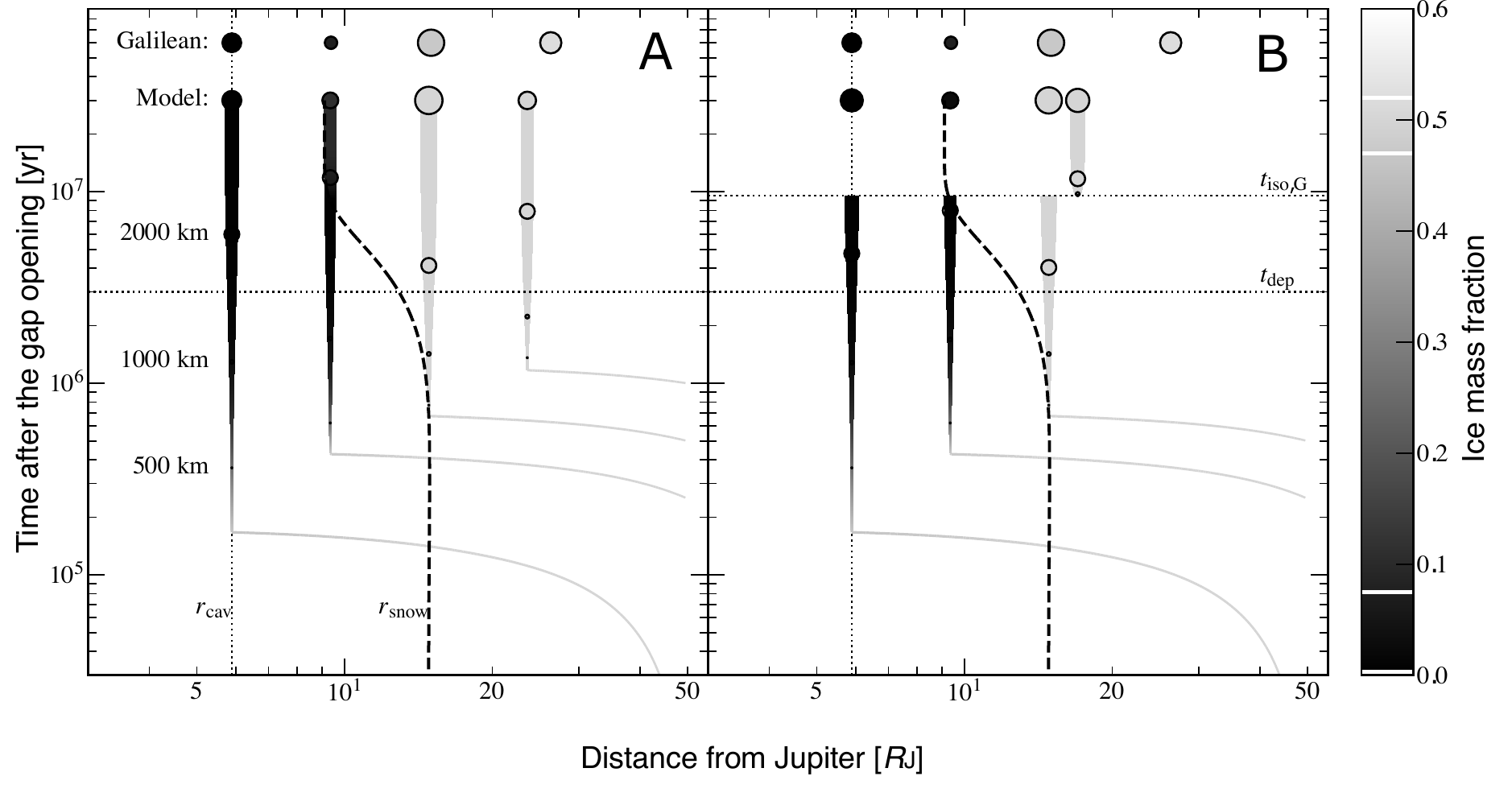}
\caption{{\bf Evolution of the Galilean satellites in the slow pebble accretion scenario.} The two panels represent the evolution in Models A and B. The sizes and the gray scales of the circles represent the mass and ice mass fractions of the satellites. The curves represent their orbital positions, sizes, and ice mass fractions. The dashed curve is the position of the snowline in this model ($r_{\rm snow}$). The vertical dotted line is the position of the edge of the inner cavity ($r_{\rm cav}$), which is assumed to be at the current orbital position of Io. The horizontal dotted lines represent the time when Ganymede reaches its pebble isolation mass ($t_{\rm PIM}$) and the depletion timescale of the disc ($\tau_{\rm dep}$). The current characteristics of the Galilean moons are also shown. The white bars in the gray scale are also the actual ice mass fractions. This figure is modified from Figures 4 and 6 of \citep{shi19}. \label{fig:pebaccretion}}
\end{figure}

\begin{figure}
\centering
\includegraphics[width=0.9\hsize]{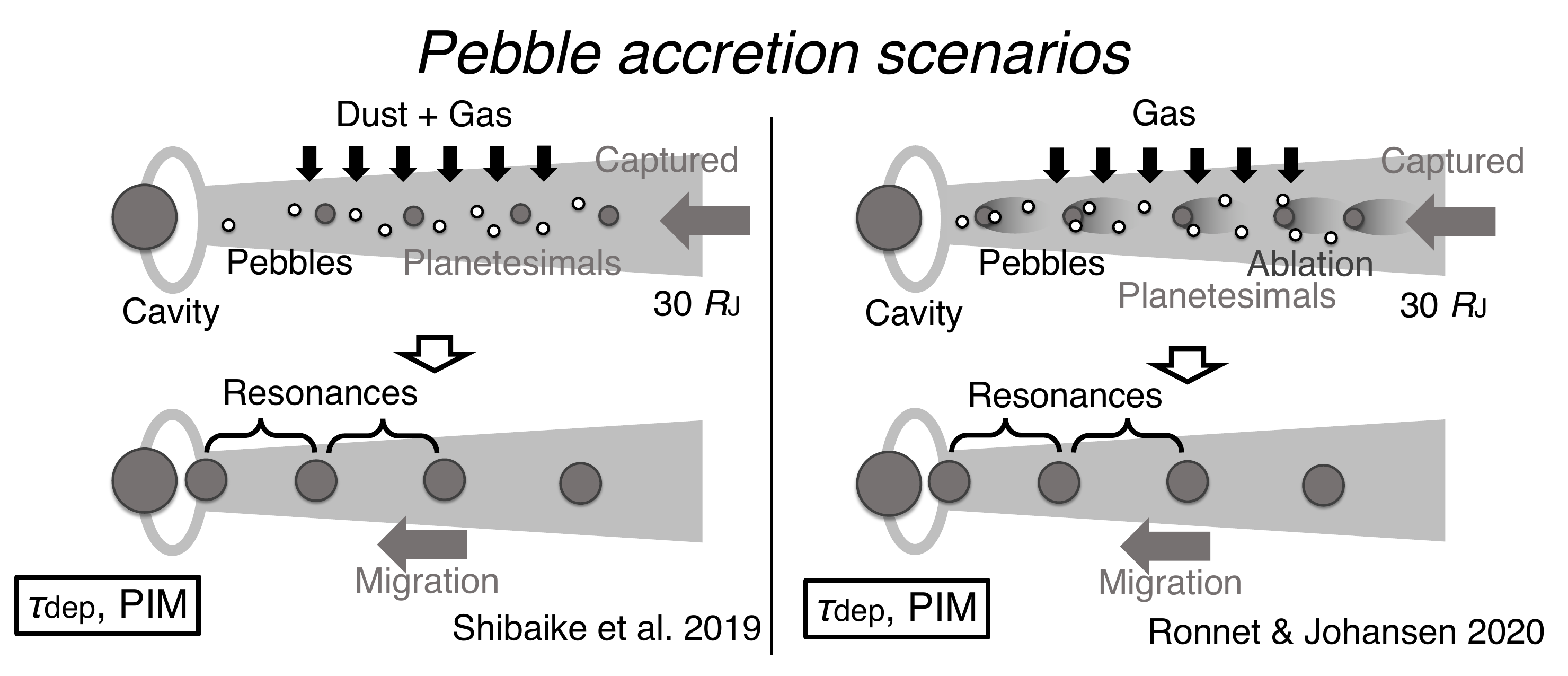}
\caption{{\bf Schematic pictures of the proposed pebble accretion scenarios.} The final mass of the moons is determined by the depletion timescale of the CJD (and CSD) and the pebble isolation mass of the satellites in both scenarios. The representative reference papers are also shown. \label{fig:pebbles}}
\end{figure}

\citet{shi19} proposed the ``slow-pebble-accretion\index{Slow-pebble-accretion}'' scenario in which relatively large planetesimals captured by the CJD accrete drifting pebbles on relatively long timescale, $\sim10^{7}~{\rm yr}$. Basically, the mass of the growing satellites can be calculated by,
\begin{equation}
M_{\rm s}(t)=\int_{t_{\rm cap}}^{t}\dot{M}_{\rm peb}P_{\rm eff}dt,
\label{Ms}
\end{equation}
where $t_{\rm cap}$, $\dot{M}_{\rm peb}$, and $P_{\rm eff}$ are the capture time of the planetesimals, the pebble mass flux in the CJD, and the pebble accretion efficiencies, respectively. From the conservation of mass, the pebble mass flux is equal to the dust accretion rate onto the CJD, $\dot{M}_{\rm peb}=\dot{M}_{\rm d}$, because a steady state is assumed. The order of magnitude of the (total) satellite-to-planet mass ratio, $M_{\rm T}/M_{\rm cp}\sim10^{-4}$ can be explained very roughly by $M_{\rm T}/M_{\rm cp}\sim\int \dot{M}_{\rm g}dt/M_{\rm cp}\times xP_{\rm eff}\sim1\times0.01\times0.01\sim10^{-4}$, when $x$ and $P_{\rm eff}$ are constant. In reality, the pebble accretion efficiency depends on the mass of the satellites and the Stokes number of the drifting pebbles \citep{liu18,orm18},
\begin{equation}
P_{\rm eff}=\left\{\left(0.32\sqrt{\dfrac{\mu_{s}\Delta v/v_{\rm K}}{{\rm St}\eta^{2}}}\right)^{-2}+\left(0.39\dfrac{\mu_{\rm s}}{\eta h_{\rm peb}}\right)^{-2} \right\}^{-1/2},
\label{Peff}
\end{equation}
where $\mu_{\rm s}=M_{\rm s}/M_{\rm cp}$ and $h_{\rm{peb}}=H_{\rm peb}/r$ are the satellite-to-central planet mass ratio and the pebble aspect ratio, respectively. The first and second terms represent the 2D and 3D regimes of pebble accretion, respectively. The relative velocity between the satellites and pebbles, $\Delta v$, is given by the Keplerian shear and the head wind,
\begin{equation}
\Delta v/v_{\rm K}=0.52(\mu_{\rm s}{\rm St})^{1/3} + \eta\left\{1 + 5.7\left(\dfrac{\mu_{\rm s}}{\eta^{3}/{\rm St}}\right)\right\}^{-1}.
\label{Deltav}
\end{equation}
Using the above equations, and the approximation of the Stokes number (an improved version of Eq. (\ref{St})), \citet{shi19} calculated the growth and Type I migration of the seeds of satellites (captured planetesimals) with the key parameters of $x=0.0026$ and $\alpha=10^{-4}$. The gas accretion rate is assumed to be $\dot{M}_{\rm g}=0.2~M_{\rm J}~{\rm Myr}^{-1}$, where $t_{\rm dep}=3~{\rm Myr}$. Here, we consider the situations that the disc has weak turbulence owing to the inactive MRI \citep{fuj14} (see Section \ref{magneticfields}). It is also assumed that three planetesimals, in other words, the seeds of Io, Europa, and Ganymede are captured by the CJD one by one and the fourth one, the seed of Callisto, is captured later (Model A) or forms at the gas pressure bump made by Ganymede (Model B). The disc inner cavity assumed to be at the current orbit of Io ($r=5.9~R_{\rm J}$) stops the migration of the innermost body, and the other three (Model A) or two (Model B) embryos are subsequently captured into the MMRs. The final mass of the satellites is regulated by the end of the supply of dust and gas except for Ganymede in Model B whose final mass is determined by its pebble isolation mass. The reasons of the slow pebble accretion in this scenario are as follows. First, the amount of dust supply is small compared to the other scenarios. Second, the dust (and gas) accretion rate onto the CJD has already decreased when the satellites grow large, resulting in the decrease of the pebble accretion efficiency. Figure \ref{fig:pebaccretion} shows the growth tracks of the satellites, which reproduce the mass and resonances of the Galilean satellites. This scenario can also explain the ice/rock mass ratio of the Galilean moons and the partial differentiation of Callisto (see next Section). The left part of Figure \ref{fig:pebbles} represents this scenario.

On the other hand, \citet{ron20} proposed another pebble accretion scenario which does not rely on the supply of dust particles; pebbles are provided by the continuous ablation of captured planetesimals. A fraction of the captured planetesimals survives and forms several proto-satellites which are large enough to start pebble accretion. It is predicted that the pebble flux inside the CJD is $\dot{M}_{\rm peb}=0.003~M_{\rm J}~{\rm Myr}^{-1}$ by this process, which is about 6 times larger than the pebble flux value assumed in \citet{shi19}. Even though the satellites grow faster, they migrate inward before they reach the current sizes of the Galilean moons, suggesting the existence of an inner cavity to stop their migration. After the innermost satellite is stopped at the inner edge of the disc, and the other three satellites are captured into the resonances, they accrete the drifting pebbles and grow at the fixed orbits. In this scenario, Ganymede reaches its pebble isolation mass after $\approx0.3~{\rm Myr}$. The right part of Figure \ref{fig:pebbles} represents this scenario.

We note that \citet{mad21} calculated N-body numerical simulations based on the pebble accretion scenarios. They found that the scattering and mutual collisions of the proto-satellites occur after they are captured into the resonance, which promotes their additional growth. This suggests that more than four seeds of satellites have to grow. In their calculations, the satellites eventually reach the multi-resonant configuration, but Callisto is also in the resonance as predicted in the other previous works.

\subsection{Constraints from observations}
\label{constraints}
As discussed in the previous section, many formation scenarios of Ganymede and the Galilean satellites have been proposed. On the other hand, the Jovian system has been the focus of many observational campaigns, from the ground, space telescopes and space probes. Here we introduce the most important characteristics of Ganymede and the Galilean moons and compare the validity of the proposed formation scenarios in light of these characteristics. We have already discussed the mass of the moons in the previous section, so that we focus on the other characteristics: resonances, ice mass fractions, and internal structures.

\subsubsection{Orbital positions and resonances}
\label{orbit}
First, the Galilean moons are in a very peculiar orbital configuration; Io-Europa and Europa-Ganymede are in the 2:1 MMRs, which constitutes a Laplace resonance\index{Laplace resonance}. However, the orbit of Callisto is not in any resonances, which implies that there are some differences in the formation process of Callisto compared to the others. There are two classical models to make the resonance chain\index{Resonance chain} system; 1) Ganymede migrates fastest because of its largest size and catches up Europa and Io as the gas disc disappears before they fall into Jupiter \citep{pea02}, and 2) the orbits of the moons evolve with differential tidal expansion by Jupiter after the gas disc disappears \citep[e.g.,][]{yod79}. \citet{pea02} showed that the former way can explain the Laplace relation of the inner three satellites better than the latter. In these two scenarios, the migration of the satellites due to the gravitational interaction with the CJD is not considered. However, as we discussed in the previous sections, the migration timescale of the satellites is much shorter than the growth timescale (and the depletion timescale of the gas disc) in any proposed scenarios. Therefore, it is natural to postulate that there is a mechanism to stop the migration during the formation process. Moreover, if the stopping occurs at the inner edge of the disc, the resonance chain of the inner three Galilean satellites can be explained naturally; the satellites form further out and migrate inward, resulting in the capture into the resonances one by one. This idea has been adopted in both satellitesimal and pebble accretion scenarios \citep{sas10,ogi12,fuj17,shi19,ron20,bat20,mad21}.

The halting mechanism is still controversial. The most investigated mechanism is the presence of a disc inner cavity\index{Inner cavity of the CJD} formed by the magnetic field of Jupiter \citep{tak96,sas10,bat18,shi19,bla25}. It is generally accepted that the magnetic field of Jupiter was stronger than today, causing magnetospheric accretion and opening a cavity, provided the gas accretion rate is small enough \citep[e.g.,][]{tak96,chr09}. The orbital position of the inner edge of the disc can be estimated by from the balance between the torque due to the Jupiter-disc magnetic interaction and the viscous torque of the disc \citep[e.g.,][]{arm10}. The position can then be around $4-5~R_{\rm J}$ during the satellite formation \citep{bat18,bat20,gin20}. Also, the edge moves outward as the gas accretion rate decreases until it reaches the corotation radius \citep{arm10}, so that Io also migrates together if the satellite has already been trapped there \citep{ogi10,liu17}. The formation and evolution theories of Jupiter show that the planet was larger, and its rotation rate was smaller then the current properties \citep[e.g.,][]{for11}. This suggests that the corotation radius was located around $r\approx5~R_{\rm J}$, which is further out than the current one ($r_{\rm co}=2.25~R_{\rm J}$) but close to the current position of Io \citep{shi19}.

Incomplete cavities can also stop the migration. Photophoresis\index{Photophoresis} is a transfer mechanism whereby gas molecules collide with photo-irradiated dust particles whose motion reflects the surrounding temperature gradient \citep[e.g.,][]{kra05}. \citet{ara19} found that the photophoresis can make a dust-poor region in the inner part of the CJD, which causes MRI only there, and the gas surface density of the inner region decreases. \citet{fuj17} found that there is an opacity transition due to the dust sublimation around the orbit of Io, which changes the radial dependence of the temperature and gas surface density distribution, and the gradients are sufficiently steep to stop the migration of the satellites.

The reason why only the orbit of Callisto is not in any resonances is an unsolved issue. If Callisto forms through the same process with the other satellites, Callisto also migrates inward and should be captured into a resonance with Ganymede. \citet{obe20} found that rapid external photoevaporation following the end of gas accretion can stop the migration of Callisto and prevent it from being captured into the resonance. Some other works mentioned the possibility that Callisto escapes from the resonance by dynamical instability after the gas disc disappeared \citep{shi19,ron20}. In the scenario by \citet{bat20}, the migration of Callisto is caused by the gravitational interaction with neighboring satellitesimals (gravitationally stirring \citep{saf69}), but the orbital excitation of the neighboring debris stops the satellitesimal-driven migration before the satellite is captured to the resonance.

\subsubsection{Ice mass fractions}
\label{icefractions}
The compositional characteristics of the Galilean moons can also constrain the formation history. The ice mass fractions of the satellites increase with their distance from Jupiter; Io is dry, Europa consist of $6-9~{\rm wt\%}$ ice, while Ganymede and Callisto have ice mass fractions of about $50~{\rm wt\%}$ \citep{kus05}. The temperature of the CJD decreases with the distance from Jupiter in any heating models, so that the variations can be explained by that qualitatively. However, it is difficult to explain that quantitatively, especially the ice mass fraction of Europa, which is rather small compared to those of the outer two moons. Some previous works based on the satellitesimal accretion scenario investigated the ice mass fractions of the satellites and found that the variation may be explained by the appropriate position and migration of the snowline \citep{can09,sas10,ogi12}. However, \citet{dwy13} found that the rocky and icy satellitesimals must be radially mixed beyond the snowline, which makes it difficult to explain the variations. On the other hand, the pebble accretion scenario should solve this issue because of the size of the icy materials; pebbles are much smaller than satellitesimals and sublimate much quicker after penetrating the inside of the snowline. \citet{ron17} found that the balance between the sublimation and drift timescales can explain the small ice fraction of Europa (with fixed snowline). \citet{shi19} also explained the small fraction by the small amount of accretion of icy pebbles due to the migration of snowline at very final phase of the formation (see Figure \ref{fig:pebaccretion}).

\citet{bie20} argued that the ice mass fractions can be explained by the hydrodynamic escape, the vapor loss from the water atmosphere, when the accretion timescale is longer than $0.3~{\rm Myr}$ in the satellitesimal accretion scenario by \citet{can02}. The warm environment at the inner part of the CJD and the deposited accretion heating\index{Accretion heating} lead to the formation of water ocean and atmosphere on the early surfaces of Io and Europa. The final density of the satellites strongly depends on the background temperature, in other words, the disc temperature at each orbital position. As a result, Io and Europa lose their water inventories completely and mostly, respectively.

We note that the temperature of the CJD does not significantly depend on the orbital position of Jupiter in the CSD \citep{hel15a}.

\subsubsection{Internal structures}
\label{internal}
The internal structures of the moons can also constraints the formation history of Ganymede and the other Galilean moons. The internal structures were inferred by the gravity measurements performed by the Galileo spacecraft and showed that the three inner moons are fully differentiated to iron cores, silicate mantles (Io), and additionally icy outer mantles (Europa and Ganymede), but Callisto is partially differentiated \citep{and01,sch04}. First, to avoid the differentiation\index{Differentiation} by the accretion heating, the formation timescale has to be long enough (longer than $0.6~{\rm Myr}$ in the case of satellitesimal accretion \citep{bar08}; see also \citet{ben25}). In most of the satellitesimal accretion scenarios, the formation timescale is $\sim0.1~{\rm Myr}$ or shorter \citep{can06,sas10,cil18}. Moreover, the dichotomy between Ganymede and Callisto requires fine-tuned conditions. Since the two satellites have similar mass and compositions, it is natural to expect that they have similar thermal and formation history \citep{bar08}. The difference of the formation timing after the calcium-aluminium-rich inclusion (CAI) formation could make the dichotomy because the $^{26}{\rm Al}$ decay radiogenic heating\index{Radiogenic heating} inside the moons depends on the accreting timing of the materials. In the slow pebble accretion scenario by \citet{shi19}, both Ganymede and Callisto accretes materials slowly enough ($\sim10~{\rm Myr}$) while Callisto forms very late, so that only the interior of Ganymede melts due to the $^{26}{\rm Al}$ decay heating. \citet{shi25b} demonstrates that Callisto’s partially differentiated interior cannot be sustained through satellitesimal accretion, but can be maintained through pebble accretion over a wide range of formation periods and onset times. In the case of pebble accretion, the accretion energy is released at the surface of Callisto, and the impact velocity is reduced by aerodynamic drag from the CPD. In the scenario by \citet{bat20}, Callisto forms after the gas disc disappears inside the satellitesimal disc with slow accretion timescale.

If both of the interiors of Ganymede and Callisto are not melted during their formation, subsequent processes may be able to melt only the interior of Ganymede: 1) the release of gravitational energy during the differentiation or 2) the Late Heavy Bombardment\index{Late Heavy Bombardment}, the concentration of impacts around $4~{\rm Gyr}$ \citep{fri83,bar10}.

We note that the ``observed'' value of the moment of inertia\index{Moment of inertia} (MOI) of Callisto is just an estimation from the in-situ gravitational field measurements assuming that Callisto is a simple hydrodynamic two-layer body. \citet{gao13} found that the estimated value of MOI of Callisto can also be reproduced even in the case that the interior is fully differentiated but has slight degree 2 distortions away from the hydrostatic shape. The JUpiter ICy moons Explorer (JUICE) mission\index{Jupiter Icy Moons Explorer (JUICE)} led by the European Space Agency (ESA) and by other future observations will test the hydro static equilibrium hypothesis and obtain more robust and precise estimates of the differentiation state \citep{cap22}. We also note that compositional gradients could cause the complete differentiation of Callisto after its formation by preventing efficient transport of radiogenic heating through convection \citep{oro14}.

\subsubsection{Toward the JUICE mission and future other observations}
As we discussed in the previous sections, many formation scenarios for Ganymede and the Galilean moons have been proposed. Each scenario has advantages and disadvantages for explaining the current characteristics of the system and, it is difficult to decide the most accurate one. One missing point which has not been frequently investigated is the history of the (icy) materials building up the moons. The compositions depend on the thermal history and the trajectory of the icy materials in the CSD and CJD, which may constrain the formation scenarios of the satellites \citep[e.g.,][]{sch25}. For example, the icy materials evaporate and then re-condense in the CJD in the classical satellitesimal accretion scenarios and in the recent pebble accretion scenario by \citet{ron20} \citep{hor08}. On the other hand, in the slow pebble accretion scenario by \citet{shi19}, the ice condenses at the outer region of the CSD, and the icy materials are transferred to the Jovian systems and accreted to the moons without evaporation through the entire journey. Thus, the icy materials might keep parts of the compositional characteristics of which have been condensed at the outer region. The characteristics are, for example, the amount of noble gas and D/H and other isotope ratios, and they depend on the surrounding environments during the condensation. The characteristics of these building blocks may have left observable signatures in the current thin atmosphere of the Galilean moons and the water plumes from Europa, which will be observed by JUICE. Moreover, if the gradient of the ice mass fractions of the Galilean moons is the result of the hydrodynamic escape (see Section \ref{icefractions}), Europa would have higher D/H and $^{18}$O/$^{16}$O ratios compared to those of Ganymede and Callisto \citep{bie20}.

\section{Conclusion}
We reviewed the origin of Ganymede and the Galilean moons in this chapter, dividing it into the three main processes: 1) the formation of the CJD, 2) the transport of the solid materials from the CSD to the CJD, and 3) the satellite formation in the CJD by the supplied materials.

The CJD is a gas disc forms around Jupiter as a byproduct of the gas accretion of the planet. Its basic structure and formation process have been revealed by both analytical modelings and hydrodynamic simulations. An envelope forms around the core of Jupiter, and then the envelope becomes a disc as the planet grows heavier. The gas accretion rate then becomes smaller as the gas surface density of the CSD decreases, which makes the CJD colder and allows the survival of the icy materials of the moons. Thus, the icy satellites should form in the later stage of the planet formation, when the gas gap has already been created by Jupiter. It has been revealed that the gas accretes to Jupiter through the CJD in spirals, but the detailed structures (such as gas surface density and direction of the gas flows) and the accretion mechanisms (such as week MRI or shock waves) are still controversial. Future numerical simulations including the effects of magnetic fields, tidal force, and photoevaporation will get more detailed characteristics of the CJD. Moreover, observations of CPDs around exoplanets have already started, which will bring us a lot of information about general characteristics of the discs.

The transport of the solid material of Ganymede and the other moons to the CJD is a newly discussed issue of their formation. The transport dominates the formation processes, but the established theory has not been developed yet. As we discussed above, when the satellite formation starts, the gas gap structure has already been formed, which prevents the transport of pebbles into the CJD. Therefore, the solid materials should be supplied as small dust particles coupled with the gas or as large planetesimals decoupled from the gas. The pebbles trapped at the edge of the gap can form planetesimals by the streaming instability and small fragments by the collision of the formed planetesimals, which may contribute to the supply of solid materials. The supply of small particles strongly depends on the gas motion, so that recent and future 3D gas+dust numerical simulations will be important. The planetesimal disc in the CSD can also present a gap structure around Jupiter but another heavy body, such as Saturn, can scatter them in the CSD, and later on the planetesimals can be captured by the CJD. The captured planetesimals can take part to the formation of the large moons in different ways: alternatives to the satellitesimals formed in-situ, seeds of the satellites, and source of dust and pebbles through the ablation.

The formation scenarios of Ganymede and the Galilean satellites have been discussed for a very long time. A lot of scenarios have been proposed, but they can be classified into the two groups following the main accretion processes of solid materials: satellitesimal accretion and pebble accretion. In the satellitesimal accretion scenarios, the fundamental issue is the formation of satellitesimals; dust particles drift inward as pebbles before they grow to satellitesimals by their mutual collisions in the CJD. Some scenarios which can avoid this drift barrier have been proposed. For example, if there is an outward flow on the mid-plane, it gathers pebbles at one place, and satellitesimals can form there by the streaming instability from the gathered pebbles. Another way is to use the  planetesimals captured by the CJD as the alternatives of the satellitesimals. In the pebble accretion scenarios, the drifting pebbles are the primary building blocks of the moons, so that the dust particles do not need to grow large to satellitesimals. Pebbles can also be created by the ablation of the captured planetesimals. The observed characteristics of the large satellites (mass, orbits, ice mass fractions, and internal structures) can contribute to constraining their formation scenarios.

In both scenarios, the migration of the satellites is important. In the classical scenarios, the final mass of the moons is determined by the balance of the growth timescale and migration timescale, which requires a large amount of continuous supplies of solid materials to the CJD. Thus, most recent scenarios consider the cases where the growth time scale is longer than the migration timescale, but the migration stops by such as the inner edge of the CJD created by the magnetic field of Jupiter. This situation is consistent with the observed resonant orbits of the inner three Galilean moons but not with the orbit of Callisto which is not in any resonances. The ice mass fractions of the moons are roughly consistent with the radial distribution of the temperature of the CJD, but the fine-tuning of the parameters is necessary in any scenarios for reproducing the characteristics quantitatively. The observed characteristics that the inner three moons are fully differentiated but Callisto is only partially differentiated also require fine-tuning in any formation scenarios. In conclusion, the decisive evidence of the one established scenario has not been found yet. Future more detailed observations by JUICE and other missions will bring us more information to find the true origin of Ganymede and the Galilean moons.

\bibliography{1-2_bibliography}

\end{document}